\documentstyle[preprint,aps]{revtex}

\begin{document}
\draft

\title{Power-law Distributions in the Kauffman Net}
\author{Amartya Bhattacharjya and Shoudan Liang}
\address{Department of Physics,
The Pennsylvania State University,
University Park, PA 16802}
\maketitle

\begin{abstract}

Kauffman net is a dynamical system of logical variables receiving two
random inputs and each randomly assigned a boolean function.
We show that the attractor and transient lengths exhibit scaleless
behavior with power-law distributions over up to ten orders of magnitude.
Our results provide evidence for the existence of the "edge of chaos" as
a distinct phase between the ordered and chaotic regimes analogous
to a critical point in statistical mechanics.
The power-law distributions are robust to the changes
in the composition of the transition rules and network dynamics.

\end{abstract}

Kauffman argues that in addition to Darwin's natural selection
and random mutation, self-organization in
random complex systems is also responsible for the observed
complexity in the biological world. A crucial question then is: what are
the characteristics of complex dynamical systems that are best able to
adapt and most suitable for evolving? It has been speculated that
such optimal systems are poised on the boundary between the ordered and
chaotic regimes\cite{PAC,LANG,kauf93}.

Random Boolean network is a dynamical system of logical variables.
Its dynamics, characterized by the period of its attractors,
tends to be chaotic at large network connectivity $k$ and ordered at
small $k$. The Kauffman net, with $k=2$, is believed to be at the edge of
chaos.
In this Letter, we study the probability distributions of both the attractor
and transient lengths of the Kauffman net. We find that the
distributions obey power-laws over up to ten orders of magnitude. The exponent
of the distribution depends on the system size, $N$, and approaches one at
large $N$. This indicates that there is no typical attractor or transient
length.
The Kauffman net therefore possesses a wide range of behaviors, which supports
the claim that it is most able to adapt\cite{kauf93,SA}.

Our finding also indicates the existence of a third phase between order and
chaos. The transition from the ordered to the chaotic phase need
not necessarily be continuous. In fact, some systems in statistical
mechanics are known to have first order transitions from an ordered to a
disordered phase. Our results provide concrete evidence for the existence
of an intermediate third regime (popularly known as the "edge of chaos") as a
distinct phase analogous to the critical point in the Ising model. Unlike in
the Ising model, the power-law behavior in the Kauffman net does not
need much special tuning. We have studied systems where the dynamics is made
more chaotic or orderly by changing the composition of the rules which
determine its dynamics. The power-law distributions remain although the
exponents are different.

The ability to study large networks of up to $N=10^5$, which is much beyond
previous works, is made possible by several algorithm innovations.
We explored a special feature of the Kauffman net in which
all the Boolean functions can be expressed in terms of logical
functions that have hardware implementation in the modern computer.
We used a method similar to ``multiple spin coding''
that packs bit variables into a machine word.
Hardware implementation of these functions enabled us to
update in parallel. At large system sizes, we used a new method to estimate
the exponents of the power-law distribution from its median value.

Over the last twenty years, random Boolean networks have been related to the
statistical properties of ontogeny and genetic regulatory
systems\cite{kauf93,kauf69}. A random Boolean network consists of a set of
binary gates $S_{i} = \{ 0,1 \}$ (called spins) interacting with
each other via certain logical rules and evolving discretely in time.
The values of the spins represent the presence and absence
of certain chemical species in a cell.
They change over time through a web of complicated chemical reactions.
The dynamics of this system of spins is as follows:
the value of spin, $S_{i}(t+1)$, at the next instant of time $t+1$ is
determined by the $k$ input spins, $S_{j_{i}^{m}}(t)(m=1,2,..k)$, at time $t$
and the transition function $f_{i}$ associated with the site $i$:
\begin{equation}
S_{i}(t+1)=
f_{i}(S_{j_{i}^{1}}(t),S_{j_{i}^{2}}(t),...,S_{j_{i}^{k}}(t))
\end{equation}

Since the chemical reactions are highly complex, as a first approximation,
the logical rules $f_{i}$ are chosen at random from amongst all the
$2^{2^{k}}$ possible Boolean rules of $k$ inputs\cite{KAUF84}. Each of the
$j_{i}^{m}$ inputs is also chosen at random from amongst the $N$ sites in the
network.
This input-output connection structure and assignment of transition
rules to each of the gates will be referred to as a {\it realization}
and a specific combination of states of the spin
as a {\it configuration} of the network. The realization is fixed in time.
Since the state space is finite and the evolution is deterministic, any
initial configuration of gates must after a finite number of time
steps re-enter a configuration it had previously encountered.
Thus, the maximum number of such time steps is $2^{N}$ and the minimum is one.
Once it revisits this configuration, the network follows the same set of
states again and does so forever. This repetitive set of states is called an
attractor or cycle of the network and its size(called the {\it cycle length}
or {\it period}) is the number of states
comprising it. The number of states from the initial configuration required
to enter an attractor is a measure of the {\it transient time} of that
configuration. Since a network realization can have many attractors of
different
sizes, the state space is divided into
these attractors by their basins of attraction which are the set
of states flowing into the attractor. Hence the attractor is an interesting
quantity to determine for the net.

In contrast to Hamiltonian systems like spin models, where the period
is at most 2\cite{deri}, random boolean nets can have any period.
Various levels of connectivity $k$ have been
analyzed. For large $k$, since the gates are randomly
connected, the system jumps randomly from one point
to another among $2^{N}$ points in the state space. The period is found to
scale
approximately as $\sqrt{2^{N}}$\cite{kauf93a}. This is the chaotic regime with
properties which are biologically impossible\cite{kauf69}.
When $k$ is reduced, the network becomes less chaotic
making a transition to orderly behavior at some small $k$.
In the annealed approximation, the transition from the
chaotic to ordered phase is found to be at $k=2$ for random
Boolean function selection\cite{derida1}. For $k=2$, known as Kauffman net,
the network flows into attractors of period much smaller than the
volume of the state space($2^{N}$).
Kauffman found the median cycle length and number of
cycles to scale approximately as $\sqrt{N}$. He related the period
to the cell replication time and the number of distinguishable
attractors to the number of different cell types in an organism with
an equivalent number of genes\cite{kauf69}.

In order to calculate the attractor and transient lengths, we use
Knuth's algorithm\cite{knuth}
with our multispin coding implementation as described at the end of the Letter.
A typical result for the period distribution is shown in Fig.\ref{rawdata}.
Although the size of the network is only 64, periods are found over a
wide range. The distribution is obtained by sampling over $5\times10^{4}$
realizations. For each realization we sample $2000$ initial
configurations which are randomly chosen amongst the $2^{64}$ possible
states. (We follow all initial configurations to its
attractor and determine the periods). The frequency of each period is
shown in Fig.\ref{rawdata}.
Notice the distinct preponderance of the even over the odd periods. The
even-odd oscillation is much bigger than the statistical noise.
We believe the even-odd
effect is an evidence for the existence of independent subnets. A subnet
may have several periods. Since the
period of a network composed of independent subnets is the least
common multiple of the periods of the subnets,
if the period of one of the subnets is even, then period
of the entire network will be even. Therefore we are more likely to find
even periods.

The second feature of the distribution is the power-law
extending over ten orders of magnitude(Fig.\ref{power-law1}). The data in
Fig.2 has been smoothened by averaging over even cycles in bins of
logarithmically increasing sizes.
The probability in a bin is obtained by summing
over the probabilities of all the periods in the bin including those not
found in the simulation. The straight lines
in Fig.\ref{power-law1} clearly show the power-law behavior with no sign of
finite size effect at large periods. The finite size deviation from the
power-law presumably occurs at very large periods of order $\sqrt{2^{N}}$.
Averaged distributions are shown for several network sizes. The
exponents of these power-laws clearly depend on the network size.

The power-law comes mostly from averaging over realizations.
Within each network realization, the distribution of attractors tend to be
clustered, {\it i.e.} the periods tends to be either all large or all small.

The exponent $\alpha$ in the distribution $f(P)=AP^{-\alpha}$ of period
$P$ is determined over the linear portion of the
averaged distribution of the even periods in Fig.\ref{power-law1}. The
even periods are used because
they give better statistics. The exponents measured with odd cycle
lengths are close to the ones from the even cycles. Another method to
determine the exponent is by computing the cumulative sum $F(Q) =
\sum_{P=1}^Q f(P)$ which smooths out the noise. The power-law
distribution for $f(P)$ implies that $(1-F(Q)) \propto Q^{1-\alpha}$ when
$\alpha> 1$. The error
on the exponents is about 3\%. The exponent for different system sizes
are listed in Table 1.

Like the attractors, we also determined the transient distributions.
The distribution has a tail of rare long transients(Fig.\ref{power-law2}). The
exponent of the power-law distribution, $\beta$, is larger than in the case of
attractors for the same $N$ and depends on $N$ significantly as shown in
Table 1.

One interesting consequence of a power-law distribution is that the mean
of the distribution diverges for $\alpha < 2$. We see from Fig.\ref{alpha}
that the exponent $\alpha$ decreases monotonically with increasing
size $N$. The exponent $\alpha$ becomes smaller than 2 for $N$ larger
than about $120$. The average of
the attractor and transient lengths determines the simulation time
and thus it becomes increasingly difficult to simulate for
large $N$. The amount of computing time is dominated by very long
cycles for large $N$ and is of the order
$(2^{N/2})^{2-\alpha}$ for $\alpha < 2$. This exponential growth with $N$
limits analysis to small $N$.

In order to study the power-law at large system size, we developed a
method that estimates the exponent of the power-law(shown in Fig.\ref{alpha}
for large $N$) from an approximate calculation of
median cycle length $P_{m}$. The median cycle length is derived
from the data of small periods up to $P_{m}$. This is very
efficient since for large $N$, $\alpha < 2$ and most of the computing time
is spent on tracking large attractors.
A configuration is evolved up to a cutoff time $M_{c}$.
The algorithm guarantees to find the transient $T$ and period
$P$ in at most $2(T+P)$ steps and we find the period of all the configurations
whose
$(T+P)$ is smaller than $M_{c}/2$. With a cutoff, the period and transient
can easily be found for a large percentage(\%s)  of initial configurations
(more than 50\%). We calculate an estimated median $\bar{P}$ from the
configurations whose periods have been determined by assuming that the rest of
the configurations have periods larger
than $\bar{P}$. The medians thus calculated are upper bounds to the
true median because an undetermined period may
be smaller than $\bar{P}$. When the cutoff $M_{c}$ is increased,
$P$ and $T$ for a larger percentage of nets are found, thus providing a
tighter upper bound. With higher cutoffs, we find that $\bar{P}$
versus \%s tends to saturate at $(60-65)\%$ for most nets. Also, the
median period is found to scale approximately as
$\sqrt N$ up to about $20000$ after which it grows faster. More details
will be available from another publication\cite{AS}. We also obtain
the probability $f(\bar{P})$ at the median. Assuming
$\bar{P}$ and $f(\bar{P})$ are very close to the true values and also
assuming that the power-law extends to the median, we can
compute the exponent from the definition of the median:
$\int ^{\infty} _{P_{m}} f(P_{m}) (\frac{P}{P_{m}}) ^{-\alpha} dP =
\frac{1}{2}$.
We obtain $\alpha -1 = 2P_{m} f(P_{m})$.  The exponent $\alpha$
determined by this method are shown as triangles in Fig.\ref{alpha}. Although
the data at large $N$ is noisy due to the difficulty associated with
determining $f(P_{m})$, the exponent agrees quite well with the direct
measurement at intermediate $N$. Our data suggest that $\alpha -1 =
C N^{-\gamma}$ where $C=5.24$, $\gamma = 0.35 \pm 0.03 $.

The power-law is robust to changes in the dynamics of the net. Updating
sequentially\cite{parseq2} produced similar power-laws with different exponents
and fewer attractors of smaller periods.
We also tried to use different combinations of transition rules.
The rules by themselves produce different dynamics.
Some of the rules are chaotic(XOR and EQUIVALENCE) in
the sense that the dynamics exclusively due to them produce enormous
periods\cite{wli111} while two others(CONTRADICTION
and TAUTOLOGY) produce a constant output of either 1 or 0.
We formed nets consisting of all except
the two chaotic or the two constant-output rules. Deleting the two chaotic
rules produces a more ordered net. Similarly,
deleting the two constant rules produces large cycles.
The distribution of cycle and transients
lengths in both cases still obey power-law with different exponents.
Our results indicate the
important contribution of the constant output rules in reducing the lengths
of the cycles as well as the number of cycles of a net.
The constant rules produce islands
of gates fixed in the evolution\cite{KAUF84}. Consequently, any cycle
consisting of these islands of frozen gates would have
fewer distinct states that can be produced by the
remaining active gates, thereby reducing the size of the cycle. Therefore,
deleting the constant rules should produce large cycle lengths.

To determine the attractors and transients, we use an
algorithm that utilizes minimum storage space\cite{knuth}.
Two identical
nets ($S(0)$ and $S'(0)$) are evolved, one at twice the speed of the other
till their Hamming distance becomes zero after say $T_{0}$ evolutions.
Then the nets $S(T_{0})$ and $S(0)$ are evolved and they become
identical after the transient length $T$. Finally $S(T)$ is evolved
and it revisits itself after $P$ steps equal to the period.
We have been able to optimize our simulation using a technique similar to
multi-spin coding in Monte Carlo simulations of Ising model\cite{stauf1}.
All the sixteen Boolean functions used was expressed as
combinations of logical functions OR, AND, EXCLUSIVE OR and NEGATIVE,
all of which have bitwise hardware implementation on a modern workstation.
Since the hardware performs logical calculation on all the bits
of a word in a CPU cycle, we can sample $N_{word}$ random initial
configurations
in parallel for a realization, where
$N_{word}$ is the size of the machine word. For each configuration assigned
to a bit, we have an individual time counter for that bit which independently
measures the time evolution along with the other bits. Whenever the period and
transient are found for a bit, a new random initial configuration is
assigned to that bit so that all the bits are always busy.

In summary, we find the probability distributions of the attractor and
transient lengths of the Kauffman net to be power-laws with the
exponent strongly dependent on the net size.
The effects of different rule composition and dynamics of the
net produce different statistics but still display power-laws.
Assuming that the Kauffman net describes adequately
the statistical properties of the genetic regulatory
network\cite{kauf93}, our result implies that the cell replication time of
organisms with the same gene size obeys a power-law.
It has been speculated that in the vicinity of phase
transitions, physical systems can behave like a computer\cite{LANG}.
Information processing arises spontaneously and is optimal in systems at
the critical state. Recently, self-organized criticality was proposed as
an underlying mechanism of certain systems to
achieve a critical state by itself (without the necessity of tuning
any external parameter) where it possesses a wide range
of behaviors described by power-law distributions\cite{BAK}.
Therefore, the power-laws\cite{stauf2} are concrete evidence
that the Kauffman net is poised in a critical state where it is most versatile.

We would like to thank Elihu Abrahams,
Wentian Li, and Leo Kadanoff for useful discussions.
We are particularly indebted to Jayanth Banavar for his constant encouragement
and contributions.
The work was supported in part by the ONR Grant No.
N00014-92-J-1340 and by a grant from the ARL at Penn State University.

\begin{table}[htbp]
\narrowtext
\begin{tabular}{||c|c|c|c|c||}
      Net Size N & $\alpha$ & $ \beta$ & $Configurations$ & $Realizations$ \\
\hline
   32 & 2.55 & 3.13& 2000 & 50000  \\
   64 & 2.24 & 2.60&2000 & 50000 \\
   128 & 1.98 & 2.01& 2000 & 50000 \\
   256 & 1.67 & 1.89 & 2000 & 3500\\
   960 & 1.49 & 2.43 &2000 & 1000  \\
\end{tabular}
\caption{Attractor and transient length distribution exponents for net size $N$
and the number of realizations and initial configurations simulated. The error
bar on $\alpha$ and $\beta$ is within $3\%$.}
\end{table}

\begin{figure}
\narrowtext
\caption{Distribution of attractors(in Log-Log scale)
for $N = 64$. The initial points are joined to
clearly show the even-odd effect with a distinct
preponderance of even over the odd cycles.}
\label{rawdata}
\bigskip

\caption{Distribution of attractors(in Log-Log scale)
for $N = 32, 128$ and $960$. The distribution is averaged over
many random realizations and configurations as noted in Table 1.
Averaging over all the periods in bins along the x axis
gives the power-law in a Log-Log plot.}
\label{power-law1}
\bigskip

\caption{Distribution of the transients(in Log-Log scale)
for $N = 32, 64, 128$. No averaging has been performed on the data.}
\label{power-law2}
\bigskip

\caption{$(\alpha - 1)$ versus $N$ for the attractor distribution,
where $\alpha$ is the attractor length exponent. Exponents for large
$N$ derived from medians are shown as triangles. A least square
fit to the data points gives slope of $-0.35 \pm 0.03$.}
\label{alpha}
\end{figure}

\end{document}